\documentclass[aps,prd,floats,floatfix,twocolumn,nofootinbib,amssymb,amsmath,showpacs]
{revtex4}
\usepackage{graphicx}
\usepackage{bm}

\newcommand{\beq}{\begin{equation}}
\newcommand{\eeq}{\end{equation}}
\newcommand{\beqa}{\begin{eqnarray}}
\newcommand{\eeqa}{\end{eqnarray}}

\usepackage[usenames]{color}
\usepackage{ulem}

\begin{document}

\title{Improved Time-Domain Accuracy Standards for Model Gravitational Waveforms}

\author{Lee Lindblom${}^1$, John G. Baker${}^2$, 
and Benjamin J. Owen${}^3$}
\affiliation{${}^1$Theoretical Astrophysics 350-17, 
California Institute of Technology, 
Pasadena, CA 91125}
\affiliation{${}^2$Gravitational Astrophysics Laboratory, 
NASA/GSFC, Greenbelt, MD 20771}
\affiliation{${}^3$Institute for Gravitation and the Cosmos,
Center for Gravitational Wave Physics,
Department of Physics,
The Pennsylvania State University,
University Park, PA 16802}

\begin{abstract}
Model gravitational waveforms must be accurate enough to be useful
  for detection of signals and measurement of their parameters, 
so appropriate accuracy standards are needed. 
  Yet these 
  standards should not be unnecessarily restrictive, making
  them impractical for the numerical and analytical modelers to meet.
  The work of Lindblom, Owen, and Brown [Phys.\ Rev.\ D {\bf 78},
    124020 (2008)] is extended by deriving new waveform accuracy
  standards which are significantly less restrictive while still
  ensuring the quality needed for gravitational-wave data analysis.  
These new standards are formulated as bounds on certain norms of the
time-domain waveform errors, which makes it possible to enforce them
in situations where frequency-domain errors may be difficult
or impossible to estimate reliably.  These standards are less
restrictive by about a factor of 20 than the previously published
time-domain standards for detection, and up to a factor of 60 for
measurement.  These new standards should therefore be much easier to
use effectively.
\end{abstract}
\date{\today}

\maketitle

\section{Introduction}
\label{s:Introduction}
Expected astrophysical sources of observable gravtiational waves
are often systems dominated by strong gravitational dynamics.  In
many cases this allows clear detailed predictions for the gravitational
waveform signals, based on Einstein's theory, General Relativity.
While these theoretical predictions are well-defined in principle,
explicit computation of the model waveforms requires a combination of
challenging analytic~\cite{Blanchet2006} and numerical~\cite{Centrella2010} 
computations to approximate the ideal theoretical expectations.  If the 
approximation errors are too large they will impact the interpretation
of gravitational wave data from operating gravitational wave instruments
such as LIGO~\cite{Abbott:2007kv} and Virgo~\cite{Acernese:2008zz}, 
or future instruments such as LISA~\cite{Jennrich:2009ti} and 
LCGT~\cite{kuroda:2010zz}.

Model waveforms are used for matched filtering in
gravitational-wave data analysis 
  both for detection and for measurement of the physical parameters of
  the source (referred to as parameter estimation by the data analysis
  community).  A signal is first identified in a detector's noisy
data stream when it is found to have a significantly large correlation
with (i.e.\ a sufficiently large projection onto) a model waveform.  If
the model waveforms were not accurate enough, then an unacceptably
large fraction of real signals would fail to be detected in this way.
The second use of model waveforms is to measure the physical
properties of any signals identified in the detection step.  These
measurements are performed by fine tuning the model-waveform
parameters (e.g., the masses and spins of the
  components of a compact binary, the source's orientation, the times
of arrival of the signals, etc.)  to achieve the largest correlation
with the data.  If the model waveforms were not accurate enough, these
measured parameters would fail to represent the true physical
properties of the sources to the level of precision commensurate with
the intrinsic quality of the data. The systematic errors due to
  modeling inaccuracies should be no greater than the statistical
  errors due to detector noise.  Model waveforms produced by
  numerical relativity simulations, e.g.\ of binary black holes, are
  computationally intensive to generate, and thus it is desirable to
  know what the minimum accuracy standards needed for these two uses
  are.  \citet{Flanagan1998} made the first derivation of abstract and
  generic accuracy standards in this context.  \citet{Miller2005} made
  the first concrete application of these standards to binary
  black-hole waveforms in numerical relativity, comparing
  post-Newtonian errors to then-current simulations and arguing for
  improvements in the latter.  Our recent work has focused on finding
  simple intuitive derivations of the fundamental accuracy standards,
  extending the standards to include the effects of detector
  calibration errors, and deriving new representations of the
  standards that are easier for the numerical relativity community to
  verify the waveform accuracy needed for gravitational-wave data
  analysis~\cite{Lindblom2008,Lindblom2009a,Lindblom2009b}.

These accuracy standards are expressed as limits on the
waveform-modeling errors $\delta h_m=h_m-h_e$, the difference between
a model waveform, $h_m$, and its exact counterpart, $h_e$.
Numerically modeled
gravitational waveforms, and the errors associated with them, are
generally evaluated most easily as functions of time, $h_m(t)$ and
$\delta h_m(t)$.  In contrast gravitational-wave data analysis and the
accuracy standards for model waveforms are most conveniently
formulated in terms of the frequency-domain representations of the
waveforms and their errors, $h_m(f)$ and $\delta h_m(f)$.  The time-
and frequency-domain representations are related to one another by
Fourier transforms, e.g.,
\begin{eqnarray}
h_m(f)=\int_{-\infty}^{\infty} h_m(t)e^{-2\pi i f t} dt.
\label{e:FourierTransform}
\end{eqnarray}
So it is straightforward (in principle) to transform from one
representation to the other.

The simplest way to express the standards needed to ensure the
appropriate levels of accuracy for model gravitational waveforms is to
write them in terms of a particular norm of the model-waveform errors:
$\langle\delta h_m|\delta h_m\rangle$.  This norm, defined by
\begin{eqnarray}
\langle \delta h_m| \delta h_m\rangle&=
&4\int_{0}^\infty \frac{
\delta h_m(f)\delta h^*_m(f)}{S_n(f)}df,
\end{eqnarray}
weights the different frequency components of the waveform error by
the power spectral density of the detector noise $S_n(f)$.  In terms
of this norm, the accuracy requirement that ensures no loss of
scientific information during the measurement process
is,
\begin{eqnarray}
\sqrt{\frac{\langle \delta h_m| \delta h_m\rangle}
{\langle h_m| h_m \rangle}} < \frac{\eta_c}{\rho},
\label{e:measurmentrealistic}
\end{eqnarray}
cf.\ Secs.~VI and VII of Ref.~\cite{Flanagan1998} and Eq.~(5) of Ref.~\cite{Lindblom2008}.  Here $\rho=\sqrt{\langle
  h_e|h_e\rangle}$ is the optimal signal-to-noise ratio of the
detected signal, and the parameter $\eta_c\lesssim 1$ is set by the
level of calibration error in the detector, cf.\
Appendix~\ref{s:CalibrationError} and Ref.~\cite{Lindblom2009a}.
Similarly the accuracy requirement that ensures no significant
reduction in the rate of detections is,
\begin{eqnarray}
\sqrt{\frac{\langle \delta h_m| \delta h_m\rangle}
{\langle h_m| h_m \rangle}} < \sqrt{2\epsilon_\mathrm{max}},
\label{e:detectideal}
\end{eqnarray}
where $\epsilon_\mathrm{max}$ is a parameter which determines the
fraction of detections that are lost due to waveform-modeling
errors, cf.\ Eq.~(14) of Ref.~\cite{Lindblom2008}. The 
choice
$\epsilon_\mathrm{max}=0.005$ ensures that no more than $10\%$ of the
signals will be missed, assuming the  template bank discretization
currently used in LIGO searches for coalescing
compact binaries~\cite{Lindblom2008}.

The basic accuracy standards, Eqs.~(\ref{e:measurmentrealistic}) and
(\ref{e:detectideal}), give the needed bounds on the waveform errors
$\delta h_m$.  Unfortunately, these abstract requirements are
difficult, or perhaps impossible, to enforce directly in
practice~\cite{Lindblom2009b}.  One fundamental problem is that an
exact knowledge of the error, $\delta h_m$, associated with a model
waveform, $h_m$, is never known.  If $\delta h_m$ were known, then the
exact waveform $h_e=h_m-\delta h_m$ would also be known, and there
would be no need for accuracy standards.  At best, the waveform
modeling community can aspire to construct tight upper bounds on these
errors, $|\delta h_m|\leq\delta H_m$, that would make it possible to
{\it guarantee} the needed waveform accuracy standards.  Such bounds
are often very difficult to construct. So in the absence of such
bounds, good estimates of the errors, $\delta h_m\approx\delta H_m$,
are a good way to verify that the accuracy standards are satisfied at
least {\it approximately}.

Another (more practical) problem with the basic accuracy standards, is
that they are formulated in terms of the frequency-domain errors
$\delta h_m(f)$.  Since numerical model waveforms are
generally computed in the time domain, $h_m(t)$, it will almost always
be easier and more straightforward to bound or estimate the
time-domain waveform errors $\delta h_m(t)$.  Time-domain quantities
are converted to the frequency-domain via the Fourier transform,
Eq.~(\ref{e:FourierTransform}).  In principle this is straightforward.
In practice, however, it is easy to introduce errors in this process
that can be larger than the intrinsic waveform modeling errors.
Discontinuities and other non-smoothness at the beginning and end of a
finite length waveform can create significant errors in the
frequency-domain representation, unless appropriate ``windowing''
procedures are followed.  This can be done
  (cf.\ Ref.~\cite{McKechan:2010kp} for a recent suggestion) but it
  requires some care and has some limitations on applicability.  Also,
  more fundamentally, a time-domain bound does not translate to a
  frequency-domain bound~\cite{Lindblom2009b}: that is, $\delta h_m(t)
  < \delta H_m(t)$ does not guarantee $\delta h_m(f) < \delta H_m(f)$.
For reasons of convenience and reliability therefore, it is very
desirable (perhaps even necessary in some cases) to have versions of
the accuracy standards formulated directly in terms of the time-domain
waveform errors $\delta h_m(t)$.  So this paper is devoted to finding
time-domain representations of these standards
that are straightforward and practical to use without being
  overly restrictive.

The most straightforward approach to obtaining the needed time-domain
expression for the waveform accuracy would be to re-write the
fundamental noise weighted norm $\langle\delta h_m|\delta h_m\rangle $
in terms of a suitably defined time-domain inner product.  This can be
done exactly 
using the ``noise kernel'' $k(t)$, defined as the real part of
the Fourier transform of $2S_n^{-1}(f)$.  The resulting time-domain
norm is then identical to the standard noise weighted frequency-domain
norm~(e.g.\ Ref.~\cite{DIS00}):
\begin{eqnarray}
\langle \delta h_m|\delta h_m\rangle &\!\!=&\!\!
\int_{-\infty}^\infty \int_{-\infty}^\infty 
\!\!\!
\delta h_m(t_1)k(t_1-t_2)\delta h_m(t_2) dt_1 dt_2.
\nonumber\\ 
\end{eqnarray}
Since this norm is identical to the standard noise-weighted norm, it
can be substituted directly into the fundamental accuracy standards,
Eqs.~(\ref{e:measurmentrealistic}) and (\ref{e:detectideal}).
Unfortunately for the same reason, it also suffers from both of the problems
discussed above.  Upper bounds in the time domain $|\delta h_m(t)|\leq
\delta H_m(t)$ do not produce upper bounds on these time-domain norms.
And errors in the norms introduced by Fourier transforms of finite
duration waveforms with non-smooth edges (Gibbs phenomena, etc.)  also
occur, because this time-domain norm is non-local 
in time on the timescales to which the detector is sensitive.  These
problems could be avoided if there were  efficient
representations of the accuracy standards that placed limits on
local-in-time norms of the waveform errors.  The purpose of this paper
is to construct new formulations of the waveform accuracy standards
that meet these criteria.

Before moving on to a detailed discussion of time-domain error
standards themselves, it is worth pausing to think briefly about how
waveform errors $\delta h_m$ have been (or could be) bounded or
estimated.  Error bounds (called a posteriori error estimates in the
mathematics literature) have been constructed for quantities derived
from the numerical solutions of hyperbolic evolution
problems~\cite{Giles.M;Suli.E2003}. However no results of this kind
are known at this time (to our knowledge) specifically for the case of
gravitational waveforms extracted from solutions of Einstein's
equations.  The error bounds that do exist are for time-domain errors,
and we know of no frequency-domain bounds for quantities derived from
the solutions of time-evolution problems at all.  Given a time-domain
error bound, $\delta H_m(t)$, it would only be useful for enforcing
the basic accuracy standards if it could be converted to a frequency
domain bound $|\delta h_m(f)|\leq \delta H_m(f)$, or if a time-domain
version of the accuracy standards were available.  Unfortunately the
Fourier transform of $\delta H_m(t)$ does not generally provide the
needed frequency-domain bound, since $ \langle \delta H_m|\delta
H_m\rangle$ is not always larger than $\langle \delta h_m|\delta
h_m\rangle$~\cite{Lindblom2009b}.  Time-domain bounds, $\delta
H_m(t)$, are only useful therefore if there are versions of the basic
accuracy standards, Eqs.~(\ref{e:measurmentrealistic}) and
(\ref{e:detectideal}), based on local-in-time norms of the waveform
error.

Good approximations of the waveform errors $\delta h_m(t)\approx
\delta H_m$, can be constructed using Richardson extrapolation
methods.  Model waveforms generally have errors that scale with some
accuracy parameter, $a$, in a well understood way.\footnote{ This
  parameter could represent the spacing between grid points for
  numerical waveform calculations, or perhaps a post-Newtonian like
  expansion parameter for an analytic calculation.} Many methods of
computing waveforms have dominant errors that scale as some power of
this parameter, $\delta h_m(t)\approx A(t) a^n$, while others have
errors that scale exponentially, $\delta h_m(t)\approx A(t)
e^{-\alpha/a}$.  In either case the values of the quantities $A(t)$,
$n$, $\alpha$, etc.\ can be evaluated by comparing the waveforms
$h_m(t,a_i)$ produced by simulations using different values of $a_i$.
The errors in the most accurate of these simulations,
$a=a_{\mathrm{min}}$, is then given approximately by the scaling
expression, e.g., $\delta h_m(t)\approx \delta H_m(t)=A(t)
a_{\mathrm{min}}^n$, using the measured values of the various
constants.  These Richardson extrapolation methods could also be
applied to frequency-domain representations of the waveforms evaluated
at different accuracies, $h_m(f,a_i)$ (subject to the problems
involved in performing the Fourier transforms discussed above).  The
resulting frequency domain error estimates, e.g., $\delta
h_m(f)\approx \delta H_m(f)= A(f) a^n_{\mathrm{min}}$, could then be
used directly to estimate whether the fundamental accuracy standards
are satisfied.  The existence of accuracy standards based on
local-in-time norms just provides a more convenient and perhaps a more
reliable way to evaluate waveform accuracy in this case, rather than
being a necessity as it was in the error-bound case.  The
gravitational waveform simulation community has begun to estimate
waveform errors by comparing time-domain waveforms computed with
different accuracy parameters~\cite{Baker2006e,Husa2007,Boyle2008a,HinderEcc2008,CampanelliEtal2009,Chu:2009md,Santamaria:2010yb,McWilliams:2010eq}.  
However, Richardson
extrapolation methods have not yet been applied to improve the quality
of these error estimates (to our knowledge).

The discussion above shows that alternate versions of the basic
accuracy standards, based on local-in-time norms of the waveform
errors, are needed for reasons of convenience and perhaps necessity.
This paper constructs a number of new versions of the standards that
meet all the needed criteria.  Some representations of the accuracy
standards based on local-in-time norms of the time-domain errors have
already been discussed in the
literature~\cite{Lindblom2008,Lindblom2009b}.  While these are
sufficient to guarantee the fundamental accuracy requirements,
Eqs.~(\ref{e:measurmentrealistic}) and (\ref{e:detectideal}), they
achieve this at the price of placing excessive restrictions on the
allowed time-domain errors.  New time-domain accuracy standards are
developed here that are considerably less restrictive: about a factor
of 20 for the detection standards, and up to a factor of 60 for
measurement.  These new standards are at most a factor of three more
restrictive than their optimal frequency-domain counterparts, so we
expect they should be practical to use.  The previous work on
time-domain accuracy standards is reviewed, and the new standards are
derived and discussed in detail in the following sections of this
paper.

\section{Time-Domain Accuracy Standards}
\label{s:TimeDomainAccuracyStandards}
The fundamental frequency-domain accuracy standards,
Eqs.~(\ref{e:measurmentrealistic}) and (\ref{e:detectideal}), can also
be expressed in terms of the time-domain $L^2$ norm of the waveform
errors~\cite{Lindblom2008,Lindblom2009b}.  These time-domain versions
provide a way to enforce the waveform-accuracy standards without the
need for a detailed knowledge of the frequency-domain representation
of the errors.  This section reviews the derivation of the
presently known time-domain standards, and illustrates the
shortcomings that make them an impractical way to enforce the accuracy
requirements.  The methods of analysis used to derive the original
time-domain standards are then generalized to produce several new and
improved versions. Each of these new standards is itself a
sufficient condition that can be used to enforce the fundamental
frequency-domain standards.  As an illustration, these new
standards are applied to the case of binary black-hole
inspiral-merger-ringdown waveforms for a detector with an
Advanced LIGO noise curve~\cite{Shoemaker2009}.  This example
shows the effectiveness and utility of these new time-domain
standards as tools for waveform-accuracy enforcement.

The frequency-domain accuracy standards, which are based on the
noise-weighted norm $\langle\delta h_m| \delta h_m\rangle$, can
be converted to time-domain standards using the following basic
inequality:
\begin{eqnarray}
\langle \delta h_m| \delta h_m\rangle&\leq
&\frac{4\int_{0}^\infty
|\delta h_m(f)|^2df}{\min[S_n(f)]}.
\label{e:BasicInequality}
\end{eqnarray}
This inequality approximates the power spectral density of the
detector noise as a constant: its minimum value.  The approximation in
Eq.~(\ref{e:BasicInequality}) is very good therefore whenever the
model waveform error is largest in the sensitive band of the
detector, and is not so good when a significant part of the error lies
outside this sensitive band.  By Parseval's theorem, the numerator in
Eq.~(\ref{e:BasicInequality}) is proportional to the time-domain $L^2$
norm, $||\delta h_m(t)||$:
\begin{eqnarray}
\!\!\!\!\!\!\!\!
||\delta h_m(t)||^2 &\equiv& \int_{-\infty}^\infty \!\!\!|\delta h_m(t)|^2 dt
=2\int_0^\infty \!\!\!|\delta h_m(f)|^2 df.
\end{eqnarray} 
Thus the basic inequality, Eq.~(\ref{e:BasicInequality}), can be
re-written as
\begin{eqnarray}
\langle \delta h_m| \delta h_m\rangle&\leq
&\frac{2||\delta h_m(t)||^2}{\min[S_n(f)]}.
\end{eqnarray}
Using this inequality, the fundamental accuracy standards,
Eqs.~(\ref{e:measurmentrealistic}) and (\ref{e:detectideal}),
can be re-written in terms of time-domain $L^2$ norms:
\begin{eqnarray}
\frac{||\delta h_m(t)||}{||h_m(t)||}\leq C_0\,\frac{\eta_c}{\rho},
\label{e:C0MeasurementStandard}
\end{eqnarray}
for measurement, and
\begin{eqnarray}
\frac{||\delta h_m(t)||}{||h_m(t)||}\leq
C_0\sqrt{2\epsilon_{\mathrm{max}}},
\label{e:C0DetectionStandard}
\end{eqnarray}
for detection.  The quantity $C_0$ that appears in these expressions
is defined as the ratio of the traditional optimal signal-to-noise
measure $\rho$ to a non-traditional measure:
\begin{eqnarray}
C^{\,2}_0 =\rho^2\left\{\frac{2||h_m(t)||^2}{\min[S_n(f)]}
\right\}^{-1}.
\label{e:C0Definition}
\end{eqnarray}
This quantity is dimensionless and independent of the overall 
scale (i.e.\ the distance to the source) used in $h_m$.

The time-domain standards, Eqs.~(\ref{e:C0MeasurementStandard})
and (\ref{e:C0DetectionStandard}), differ therefore from the
fundamental standards, Eqs.~(\ref{e:measurmentrealistic}) and
(\ref{e:detectideal}), in just two ways: {\it i)} the norms used to
measure the waveform errors on the left sides are different, and {\it
  ii)} the maximum allowed errors on the right sides are multiplied by
the factor $C_0$ in the time-domain case.  It is straightforward to
show that $C_0$ is always less than one: $C_0\leq 1$.  Thus the
time-domain standards are always more restrictive than their
frequency-domain counterparts.  The quantity $C_0$ compensates in the
time-domain standards for the fact that the waveform errors are not
being weighted by the detector noise in the optimal way.
Unfortunately these time-domain standards have two serious flaws: The
first is that the time-domain norms $||h_e(t)||$ are not well defined,
because the waveforms $h_e(t)$ do not fall to zero quickly enough as
$t\rightarrow -\infty$.  Since most model waveforms $h_m(t)$ are only
evaluated for finite time intervals, this problem is one of principle
more than practice.  But this does mean that the norms $||h_m(t)||$,
and thus the quantity $C_0$, depend (although fairly weakly) on the
length in time of the model waveform.  The second flaw, discussed in
more detail below, is that $C_0$ is quite small for realistic
waveforms, thus making these time-domain versions of the standards
very over-restrictive. 
\begin{figure}[t]
\centerline{\includegraphics[width=3in]{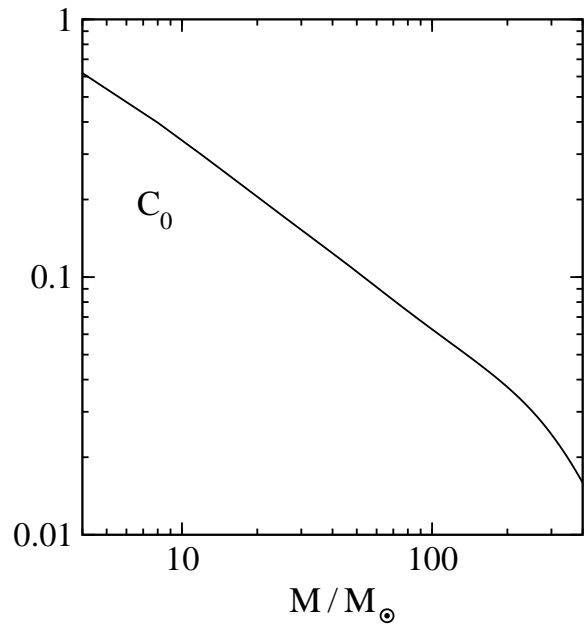}}
\caption{\label{f:C0Plot} Curve illustrates $C_0$, the ratio of the
  traditional optimal signal-to-noise measure $\rho$ to a
  non-traditional measure defined in Eq.~(\ref{e:C0Definition}), as a
  function of the total mass for non-spinning equal-mass black-hole
  binary waveforms and for a detector with an Advanced LIGO noise
  curve optimized for double neutron star binaries.}
\end{figure}

To determine how overly restrictive the time-domain standards are in
practice, the quantity $C_0$ has been evaluated for non-spinning
equal-mass black-hole binary waveforms and for a detector with an
Advanced LIGO noise curve optimized for double neutron star
binaries~\cite{Shoemaker2009}.\footnote{The data for this neutron-star
    optimized noise
    curve comes from the file NSNS\_Opt.txt available from the URL in
    Ref.~\cite{Shoemaker2009}.}
These waveforms were constructed by patching together post-Newtonian
waveforms for the early inspiral phase, with numerical relativity
waveforms for the late inspiral plus merger plus ringdown
phases~\cite{Scheel2008,Boyle2008b} and adjusting the time scales for
different total masses.  They are the same used earlier to evaluate
$C_0$~\cite{Lindblom2008}.  Figure~\ref{f:C0Plot} illustrates $C_0$ as
a function of mass for binary black-hole systems with masses in the
range that is relevant for Advanced LIGO: $4\leq M/M_\odot\leq 400$.
For low mass binary systems---where these model waveforms lie mostly
within the Advanced LIGO frequency band---$C_0$ is of order unity, and
the time-domain standards are reasonably efficient representations of
the fundamental frequency-domain standards.  However, for binary
systems with large masses---where these model waveforms lie mostly
outside the Advanced LIGO band---$C_0$ becomes very small, and the
time-domain standards become very ineffective.  For the largest masses
the quantity $C_0$ becomes very small, $C_0\approx 0.016$, and the
time-domain standards must place unreasonably tight constraints on the
waveform errors to ensure that the fundamental frequency-domain
standards are satisfied by the small part of these waveforms lying
within the Advanced LIGO band.

The methods of analysis used to derive
Eqs.~(\ref{e:C0MeasurementStandard}) and (\ref{e:C0DetectionStandard})
can be generalized to obtain a new set of time-domain accuracy
standards.  This can be done using the following extension of the
basic inequality:
\begin{eqnarray}
\langle \delta h_m| \delta h_m\rangle&\leq
&\frac{4\int_{0}^\infty(2\pi f)^{2k} |\delta h_m(f)|^2df}{\min[(2\pi
    f)^{2k}S_n(f)]}.
\label{e:BasicKInequality}
\end{eqnarray}
The numerator in this expression is just the frequency-domain $L^2$
norm of the $k^{\mathrm{th}}$ time derivative of the waveform error.
This can be converted to a time-domain $L^2$ norm using Parseval's
theorem:
\begin{eqnarray}
||\delta h_m(t)||_k^2 &\equiv&
\int_{-\infty}^\infty \left|\frac{d^k \delta h_m(t)}{dt^k}\right|^2 dt,
\nonumber\\
&=& 2\int_0^\infty(2\pi f)^{2k} |\delta h_m(f)|^2df.
\label{e:kNormDefinition}
\end{eqnarray}
The basic inequality in Eq.~(\ref{e:BasicKInequality}) can therefore
be re-written as
\begin{eqnarray}
\langle \delta h_m| \delta h_m\rangle&\leq
&\frac{2||\delta h_m(t)||_k^2}{\min[(2\pi f)^{2k}S_n(f)]}.
\end{eqnarray}
These inequalities together with the fundamental frequency-domain
accuracy standards, Eqs.~(\ref{e:measurmentrealistic}) and
(\ref{e:detectideal}), can then be used to obtain a new set of
time-domain accuracy standards:
\begin{eqnarray}
{\cal E}_k
\leq C_k\,\frac{\eta_c}{\rho},
\label{e:CkMeasurementStandard}
\end{eqnarray}
for measurement, and
\begin{eqnarray}
{\cal E}_k\leq
C_k\sqrt{2\epsilon_{\mathrm{max}}},
\label{e:CkDetectionStandard}
\end{eqnarray}
for detection. The quantities ${\cal E}_k$,
\begin{eqnarray}
{\cal E}_k \equiv
\frac{||\delta h_m(t)||_k}{||h_m(t)||_k},
\label{e:ErrorKDefinition}
\end{eqnarray}
are new measures of the time-domain waveform error, and the quantities
$C_k$ are defined as the ratios of the traditional optimal
signal-to-noise measure $\rho$ to new non-traditional measures:
\begin{eqnarray}
C_k^{\,2} =\rho^2\left\{\frac{2||h_m(t)||^2_k}{\min[(2\pi f)^{2k}S_n(f)]}
\right\}^{-1}.
\label{e:CkDefinition}
\end{eqnarray}
Both quantities, ${\cal E}_k$ and $C_k$, are dimensionless and
independent of the overall scale (i.e. the distance to the source)
used in $h_m$.  It is straightforward to show that $C_k\leq 1$, so
these new standards are always more restrictive than their
frequency-domain counterparts.  The $k=0$ versions of
Eqs.~(\ref{e:CkMeasurementStandard}) and (\ref{e:CkDetectionStandard})
are identical to the previously known time-domain standards,
Eqs.~(\ref{e:C0MeasurementStandard}) and
(\ref{e:C0DetectionStandard}).  But the standards with $k\geq 1$ are
new and distinct.  The new standards with $k\geq 1$ are free from the
flaws described earlier for the previously known $k=0$ standards.  The
norms $||h_m(t)||_k$ are well defined for $k\geq1$, because 
$d^k h_m/dt^k$ falls 
off quickly enough as $t\rightarrow -\infty$.  And as
described in more detail below, the $C_k$ for $k\geq 1$ are much
closer to one.  So these new standards are much more effective ways to
enforce the accuracy requirements.

\begin{figure}[t]
\centerline{\includegraphics[width=3in]{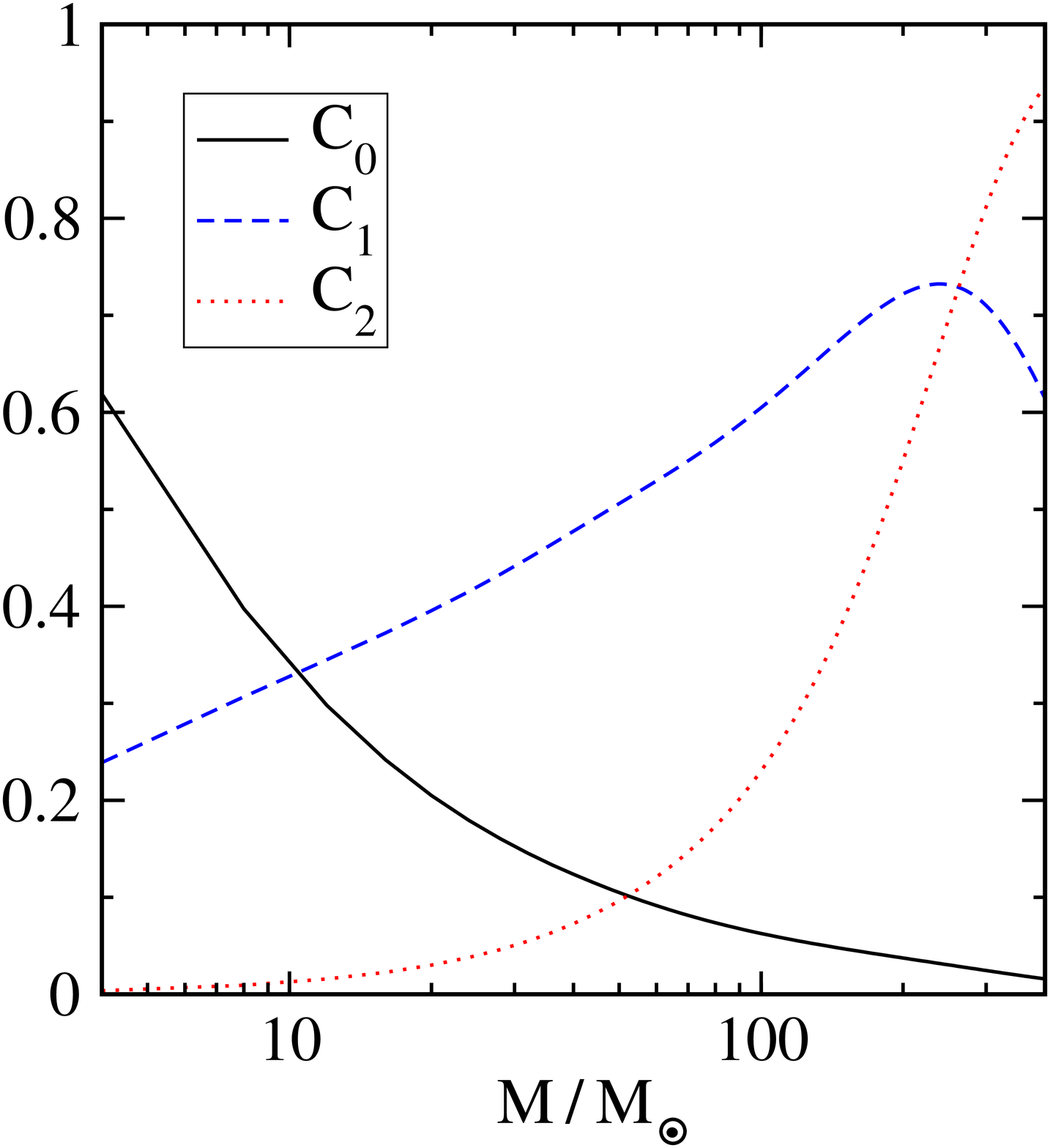}}
\caption{\label{f:CkPlot} Curves illustrate $C_k$, the ratio of the
  traditional 
  optimal signal-to-noise measure $\rho$ to a
  non-traditional 
  measure defined in Eq.~(\ref{e:CkDefinition}), as a
  function of the total mass for non-spinning equal-mass black-hole
  binary waveforms.  The waveform accuracy standards,
  Eqs.~(\ref{e:CkMeasurementStandard})--(\ref{e:CkDetectionStandard}),
  can be enforced using any value of $k$.  Thus it is sufficient to
  impose the standard having the largest $C_k$ in each mass range.}
\end{figure}
Figure~\ref{f:CkPlot} illustrates the quantities $C_k$ (for $k=0, 1$,
and 2) that appear in the waveform accuracy standards,
Eqs.~(\ref{e:CkMeasurementStandard}) and
(\ref{e:CkDetectionStandard}), as a function of the mass of the
waveform.  Each of these new accuracy standards is sufficient to
guarantee that the corresponding frequency-domain standard is
satisfied.  Choosing the easiest standard to satisfy is all that is
required for a given $M$, and this will typically be the one
having the largest $C_k$.  Figure~\ref{f:CkPlot} shows that the
standards with different values of $k$ have different mass ranges
where they are most effective.  The $k=0$ standard is most effective
for low-mass binary waveforms, while the $k=2$ standard is the most
effective for high-mass waveforms.

\begin{figure}[t]
\centerline{\includegraphics[width=3in]{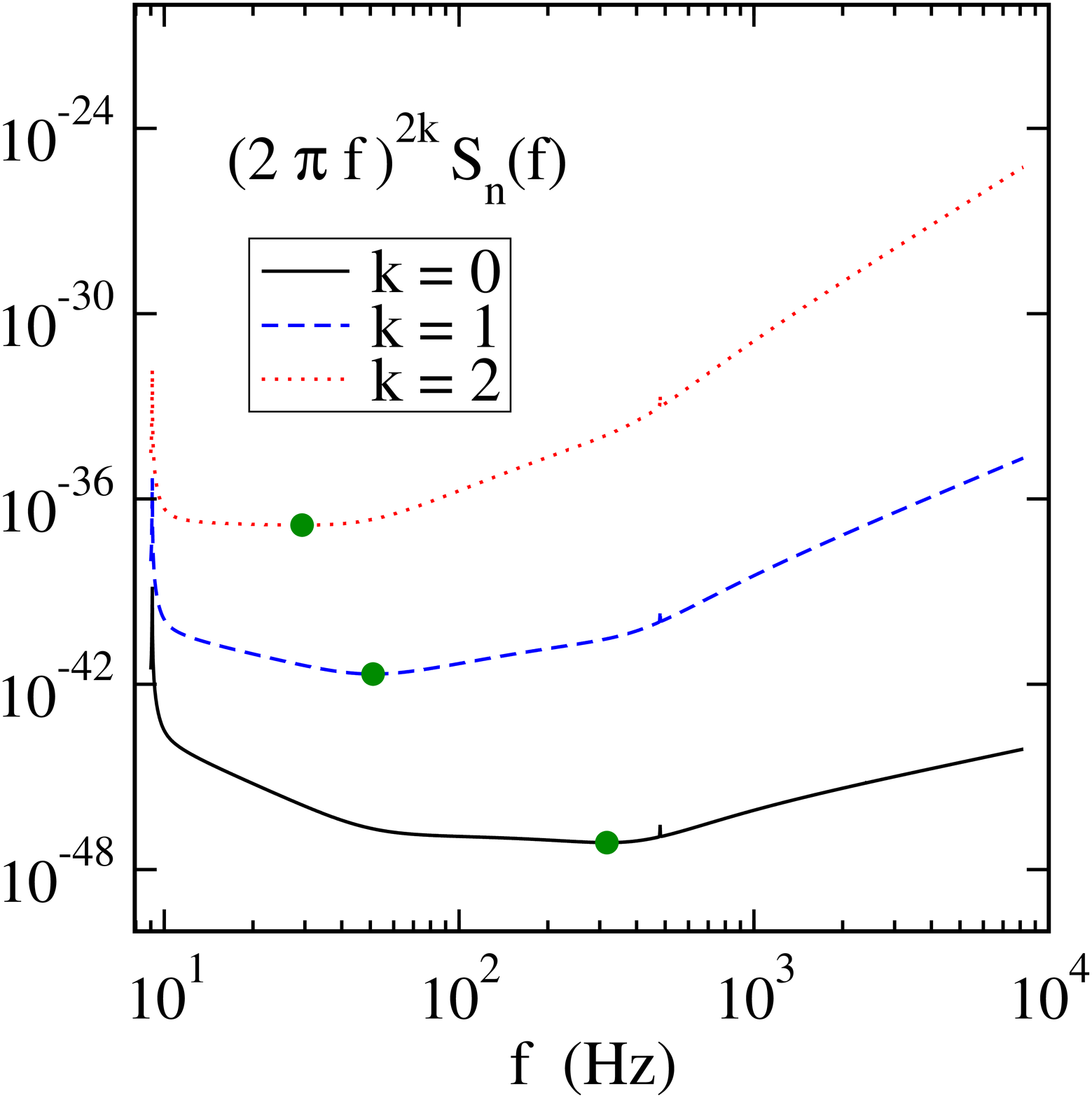}}
\caption{\label{f:NoiseCurves} Curves illustrate the effective noise
  curves, $(2\pi f)^{2k} S_n(f)$, that appear in the norms of the
  $k^{\mathrm{th}}$ time derivatives of the waveform.  The minima of
  these curves are marked by the large dots.  These curves illustrate
  why the higher time derivative norms are more sensitive to the lower
  frequency parts of the waveforms, and hence do a better job of
  modeling the waveform accuracy in high mass binaries.}
\end{figure}

Figure~\ref{f:NoiseCurves} shows the noise curves used to compute the
$C_k$ illustrated in Fig.~\ref{f:CkPlot}. These particular noise
curves are all representations of the Advanced LIGO sensitivity curves
optimized for neutron-star/neutron-star binary
signals~\cite{Shoemaker2009}.  The noise curves with successively
larger values of $k$ have minima at successively smaller values of the
frequency $f$.  Figure~\ref{f:NoiseCurves} shows that the factor
$f^{2k}$ combines with the piecewise power law structure of $S_n(f)$
to produce fairly broad regions where the effective noise curves are
rather flat.  In these flat regions the approximation,
$\min[f^{2k}S_n(f)]\approx f^{2k}S_n(f)$, becomes extremely good, and
the approximation leading to Eq.~(\ref{e:BasicKInequality}) becomes
nearly exact.  These features explain one of the reasons the higher
time-derivative norms measure the lower-frequency parts of the
waveforms more faithfully, and hence do a better job of enforcing the
waveform accuracy standards in high-mass waveforms.  The higher
time-derivative norms measure $d^k h/dt^k$, which is more strongly
peaked near the time of merger.  Hence the energy norms ${\cal E}_k$
for $k\geq 1$ do a better job than ${\cal E}_0$ of measuring the
strongest parts of the waveform that contribute most to detection.

Model waveforms used to measure accurately the physical parameters of
a previously detected signal may be custom made for this
task.  The physical parameters of the waveform will be known
(approximately) in this case from the detection process, so a finely
targeted template bank of waveforms with parameters that cover a small
neighborhood surrounding the actual physical values can be computed.
For such applications, the quantities $C_k$ can easily be computed
along with the model waveforms.  The waveform accuracy standard for
measurement, Eq.~(\ref{e:CkMeasurementStandard}), should be enforced
therefore using the standard having the largest $C_k$ for these
particular waveforms.  As Fig.~\ref{f:CkPlot} shows, this will
typically be the $k=2$ standard for the largest mass binary waveforms,
$k=0$ for the smallest mass waveforms, and $k=1$ for a broad range of
intermediate mass waveforms.  The smallest value of $C_k$ needed for
these measurement tests is $C_0=C_1\approx 0.331$ if the mass of the
waveform is near $10M_\odot$.  However, if the mass is larger, the
value of the $C_k$ could be as large as $C_2\approx 0.935$ for
waveforms with masses near $400M_\odot$.  These new time-domain
accuracy standards for measurement can therefore be up to 60 times
less restrictive (depending on the mass of the waveform) than the
previously available $k=0$ standard~\cite{Lindblom2008}.

The situation is somewhat more complicated for waveforms used as part
of the detection process.  These waveforms should be scalable to the
full range of masses whose waveforms lie within the sensitivity band
of the detector.  The simplest way to do this would be to use the
accuracy standard having the largest $\min_{M} C_k(M)$, where the
minimization is done over the waveform masses in the appropriate
range.  Note that the quantity ${\cal E}_k$, which appears on the left
side of Eq.~(\ref{e:CkDetectionStandard}), is independent of mass
because both $||\delta h_m||_k$ and $||h_m||_k$ scale with mass in
exactly the same way.  Thus the accuracy standard for a given $k$ can
be enforced over the entire range of masses by ensuring that ${\cal
  E}_k$ satisfies the standard with $C_k=\min_{M} C_k(M)$.  The $k=1$
standard has the largest $\min_{M} C_k(M)$ for the waveforms
illustrated in Fig.~\ref{f:CkPlot}, so it is sufficient in this case
to enforce the $k=1$ standard with $C_1=\min_{M} C_1(M)\approx 0.239$.
This is about 15 times less restrictive than would be required using
the previously available $k=0$ standard where $\min_{M} C_0(M)\approx
0.016$.

A less restrictive accuracy standard for detection can also be
constructed, at the expense of making it slightly more complicated.
Figure~\ref{f:CkPlot} shows that enforcing the detection standard,
Eq.~(\ref{e:CkDetectionStandard}), with $C_1=\min_M C_1(M)$ is
somewhat more restrictive than is really necessary.  All that is
required is to enforce the standard having the largest $C_k(M)$ for
each mass.  So it would be sufficient to enforce the $k=0$ standard in
the mass range where $C_0$ is the largest of the $C_k$: $M\leq
M_{\mathrm{int}}$.  The mass $M_{\mathrm{int}}\approx 10.4M_\odot$
represents the point where the $C_0(M)$ and the $C_1(M)$ curves
intersect: $C_0(M_{\mathrm{int}})=C_1(M_{\mathrm{int}}) \approx 0.331$
for the waveform example shown in Fig.~\ref{f:CkPlot}.  It would also
be sufficient to enforce the $k=1$ standard in the mass range $M \geq
M_{\mathrm{int}}$.  Therefore it is sufficient over the full range of
masses to enforce both the $k=0$ and the $k=1$ versions of the
detection accuracy standard, Eq.~(\ref{e:CkDetectionStandard}), with
$C_0(M_{\mathrm{int}})=C_1(M_{\mathrm{int}})\approx
0.331$:\footnote{In principle it would be sufficient to enforce the
  $k=2$ instead of the $k=1$ standard in the mass range $M\gtrsim
  260M_\odot$.  However, this would not make the needed $k=1$
  condition any weaker, because ${\cal E}_1$ is independent of mass.}
\begin{eqnarray}
{\cal E}_0\lesssim 0.331\,\sqrt{2\epsilon_{\max}},\quad
\mathrm{and} \quad
{\cal E}_1\lesssim 0.331\,\sqrt{2\epsilon_{\max}}.
\label{e:C0C1DetectionStandard}
\end{eqnarray}
This condition is about 20 times less restrictive than would be
required to enforce the previously available $k=0$ standard over the
full range of masses. The overall factor that appears on the
  right sides in Eq.~(\ref{e:C0C1DetectionStandard}) appears to be
  quite insensitive to the detector noise curve, giving 0.351 instead
  of 0.331 using the ``zero detuning, high-power'' version of the
  Advanced LIGO noise curve.\footnote{The data for this zero-detuning,
  high power noise curve comes from the
  file ZERO\_DET\_high\_P.txt available from the URL 
  in Ref.~\cite{Shoemaker2009}.}  
  We have not yet
  determined how sensitive this factor is to properties of the
  waveform models, like the total duration of the waveform, or the
  spins of the individual black holes.  

\section{Discussion}%
\label{s:Discussion}

New time-domain representations of the waveform accuracy
standards for gravitational-wave data analysis have been presented in
Sec.~\ref{s:TimeDomainAccuracyStandards}.  For waveforms to be used
for measuring the physical properties of a detected signal, the new
time-domain standards, Eq.~(\ref{e:CkMeasurementStandard}), provide
the best way to enforce the needed accuracy standards.  These
standards give the allowed upper limits on certain new measures of the
waveform error ${\cal E}_k$, defined in terms of the time-domain $L^2$
norms of the $k^{\mathrm{th}}$ time derivatives of the model waveform
$h_m$ and its error $\delta h_m$: ${\cal E}_k\equiv||\delta
h_m(t)||_k/||h_m(t)||_k$. Model waveforms are generally obtained by
performing time integrals of an equation like $d^2 h_m(t)/dt^2
= \tilde\Psi_4(t)$, where $\tilde\Psi_4(t)$ is the (real) projection
of the outgoing component of the Weyl curvature representing
  the wave polarization of interest.  Therefore, it is no harder or
less accurate to evaluate the first two time derivatives of $h_m(t)$,
and their accuracy, than it is to evaluate $h_m(t)$ and its accuracy.
For the binary black-hole waveforms used as an example here; the $k=0$
version of this standard is most effective for black-hole systems with
masses below about $10M_\odot$; the $k=1$ standard is best in the
intermediate mass range $10\lesssim M/M_\odot\lesssim 260$; and the
$k=2$ standard is best for the highest mass waveforms, $260\lesssim
M/M_\odot\leq 400$.  This new version of the accuracy standard for
measurement is less restrictive than the previously available
time-domain standard by up to a factor of 60 for the largest mass
binary systems, and is never more restrictive than the
ideal frequency-domain standards by a factor
greater than about 3.  The new standard for measurement is
nevertheless sufficient to ensure that the 
  ideal frequency-domain standard is satisfied as well.

The situation is slightly more complicated for the new time-domain
waveform accuracy standards for detection.  These standards put limits
on both the $k=0$ and $k=1$ versions of the time-domain error
measures, ${\cal E}_0$ and ${\cal E}_1$. The best new time-domain
standard for detection presented in
Sec.~\ref{s:TimeDomainAccuracyStandards} enforces these standards with
$C_0=C_1\approx 0.331$ for the example waveform used here.  This new
standard for detection, Eq.~(\ref{e:C0C1DetectionStandard}),
guarantees that the frequency-domain standards are enforced, and it is
only about a factor of three more restrictive than those  ideal
standards.  The new standards should be effective enough therefore to
make them usable without placing an undue burden on the
waveform-simulation community.  Slightly less restrictive time-domain
standards for detections can also be derived using methods similar to
those used in Sec.~\ref{s:TimeDomainAccuracyStandards}.  The
derivation of these additional alternative standards is discussed and
analyzed in detail in Appendix~\ref{s:CompoundStandards}.  These 
alternative standards are more complicated, however, and consequently
may not prove to be as useful as Eq.~(\ref{e:C0C1DetectionStandard}).

\acknowledgments We thank Michael Holst for valuable discussions about
rigorous mathematical error bounds, Sean McWilliams for other
useful discussions, and Duncan Brown for helpful comments on an earlier
draft of this paper.  This research was supported in part by grants to
Caltech from the Sherman Fairchild Foundation, NSF grants DMS-0553302,
PHY-0601459, and PHY-0652995, and NASA grant NNX09AF97G; by NASA
grants 08-ATFP08-0126 and 09-ATP09-0136 to Goddard Space Flight
Center; by NSF grant PHY-0855589 to Penn State; and by the LIGO
Visitors Program.  LIGO was constructed by the California Institute of
Technology and Massachusetts Institute of Technology with funding from
the National Science Foundation and operates under cooperative
agreement PHY-0757058.  This paper has LIGO Document Number
LIGO-P1000078-v2.


\appendix
\section{Calibration Error Levels}
\label{s:CalibrationError}

The basic accuracy requirement for measurement,
Eq.~(\ref{e:measurmentrealistic}), reduces to the case of an ideal
detector when $\eta_c=1$.  The ideal detector approximation means that
the systematic errors made in calibrating the detector (by measuring the response
function of the detector) are negligible compared to the statistical
errors due to the intrinsic
noise level of the detector.  It is more realistic to expect that in
many cases the detector will have non-negligible calibration errors.
These errors influence the data analysis process in almost the same
way as waveform modeling errors~\cite{Lindblom2009a}.  Scientific
information will be lost unless the combined systematic calibration and waveform
modeling errors are smaller than the intrinsic detector noise
level.  This requires that both the calibration error and the waveform
modeling errors be kept below the intrinsic noise level of the
detector.  The allowed error budget may however be apportioned
arbitrarily between the two sources, e.g., in the most economical or
convenient way.  The parameter $\eta_c$ introduced in
Eq.~(\ref{e:measurmentrealistic}) determines the share of this error
budget allowed for waveform modeling error.  In particular it
represents the amount by which the waveform modeling error must be reduced
below the intrinsic detector noise level of the detector to allow for
the presence of calibration error.

In some cases detectors may have calibration errors that are
negligible compared to the needed waveform modeling errors (e.g., the
proposed Laser Interferometer Space Antenna LISA).  In those cases the
ideal-detector conditions apply and $\eta_c=1$ is the appropriate
choice.  In most cases, however, it may only be possible or
economically feasible to calibrate to an accuracy level comparable to
the intrinsic detector noise.  In these cases it seems likely that it
will be optimal to split the available error budget equally between
the calibration and waveform-modeling errors.  If these errors can be
considered uncorrelated, then $\eta_c=1/\sqrt{2}$ is the appropriate
choice.  Waveform-modeling errors, $\delta h_m$, are uncorrelated with
the calibration errors, $\delta h_R$, if $\langle \delta h_m|\delta
h_R\rangle^2 \ll \langle \delta h_m|\delta h_m\rangle \langle \delta
h_R|\delta h_R\rangle $.  To justify taking $\eta_c=1/\sqrt{2}$, this
uncorrelated condition should be satisfied for all times when data is
being collected, and for all the different errors associated with the
various model waveforms that will be used in the subsequent data
analysis.  While calibration and modeling errors come from completely
unrelated sources, it may be difficult to completely rule out the
possibility of significant correlations.  For example, it is easy to
imagine both calibration and modeling errors that are proportional to
the exact waveform $h_e$.  When correlations can not be ruled out, the
somewhat stricter condition, $\eta_c=\frac{1}{2}$, is required.  The
calibration accuracy needed to ensure no loss of scientific
information for the strongest sources is generally much greater than
the accuracy needed for detection.  The weak dependence of the
detection standard on $\eta_c$ has therefore been ignored here, and
the ideal-detector version of the standards,
Eq.~(\ref{e:detectideal}), are used in this paper.

\section{Compound Accuracy Standards}
\label{s:CompoundStandards}

A somewhat more general class of 
time-domain accuracy standards
are derived in this Appendix using the methods developed in
Sec.~\ref{s:TimeDomainAccuracyStandards}.  These additional standards
are somewhat less restrictive for detection than
Eq.~(\ref{e:C0C1DetectionStandard}), but they are also 
more complex.  The idea is to construct a new measure of the waveform
error that uses a combination of the time-domain norms used in
Eq.~(\ref{e:CkDetectionStandard}).  This is done by splitting the
integral that appears in the basic inequality,
Eq.~(\ref{e:BasicKInequality}), into parts proportional to $\mu$ and
$1-\mu$ respectively (where $\mu$ 
is an arbitrary splitting factor that satisfies $0\leq\mu\leq 1$),
and then applying the constant-noise approximation using different
values of $k$ for each part.  The result is a compound version of the
basic inequality:
\begin{eqnarray}
\!\!\!\!\!\!\!\!\!\!\!\!
&&\langle \delta h_m| \delta h_m\rangle\leq
\frac{4\mu\int_{0}^\infty(2\pi f)^{2k} |\delta h_m(f)|^2df}{\min[(2\pi
    f)^{2k}S_n(f)]}\qquad\qquad\quad\nonumber\\
&&\qquad\qquad\quad
+\frac{4(1-\mu)\int_{0}^\infty(2\pi f)^{2k'} |\delta h_m(f)|^2df}{\min[(2\pi
    f)^{2k'}S_n(f)]}.
\label{e:BasicKCompoundInequality}
\end{eqnarray}
The right side of this inequality can be re-written in terms of the
time-domain error measures ${\cal E}_k$ of
Eq.~(\ref{e:ErrorKDefinition}), and the quantities $C_k$ of
Eq.~(\ref{e:CkDefinition}):
\begin{eqnarray}
\langle \delta h_m| \delta h_m\rangle&\leq
&\frac{\mu\,{\cal E}_k^2}{C_k^2}
+\frac{(1-\mu)\,{\cal E}_{k'}^2}{C_{k'}^2}.
\end{eqnarray}
This basic inequality can be transformed into a more compact and
useful form by replacing the splitting parameter $\mu$ with a
new parameter $\sigma$, defined by
\begin{eqnarray}
\mu=\frac{\sigma C^2_{k}}{\sigma C^2_{k}+(1-\sigma)C^2_{k'}} .
\end{eqnarray}
The parameter $\mu$ satisfies the required inequality,
$0\leq\mu\leq 1$, whenever $0\leq \sigma\leq 1$.
The basic inequality then becomes,
\begin{eqnarray}
\langle \delta h_m| \delta h_m\rangle&\leq&
\frac{{\cal E}^2_{k,k'}(\sigma)}{C^2_{k,k'}(\sigma)},
\label{e:CompBasicInequality}
\end{eqnarray}
where ${\cal E}_{k,k'}(\sigma)$ is a composite measure of the
waveform error, defined by
\begin{eqnarray}
{\cal E}^2_{k,k'}(\sigma)=\sigma {\cal E}_k^2 +(1-\sigma){\cal
  E}_{k'}^2,
\label{e:CompoundErrorNorm}
\end{eqnarray}
and $C_{k,k'}(\sigma)$ is the composite analog of $C_k$, defined by
\begin{eqnarray}
C^2_{k,k'}(\sigma)=\sigma C^2_k + (1-\sigma)C^2_{k'}.
\end{eqnarray}
This new form of the basic inequality,
Eq.~(\ref{e:CompBasicInequality}), is true for any $\sigma$.  It can
be used to convert the fundamental frequency-domain standards,
Eqs.~(\ref{e:measurmentrealistic}) and (\ref{e:detectideal}), into new
time-domain waveform accuracy standards:
\begin{eqnarray}
{\cal E}_{k,k'}(\sigma)\leq 
  C_{k,k'}(\sigma)\frac{\eta_c}{\rho},
\label{e:CSigmaMeasurementStandard}
\end{eqnarray}
for measurement, and
\begin{eqnarray}
{\cal E}_{k,k'}(\sigma)\leq
C_{k,k'}(\sigma)\sqrt{2\epsilon_{\mathrm{max}}},
\label{e:CSigmaDetectionStandard}
\end{eqnarray}
for detection.  The waveform error measure ${\cal E}_{k,k'}(\sigma)$
is positive definite for all $\sigma$ in the range $0\leq\sigma\leq
1$.  So these new time-domain waveform accuracy standards,
Eqs.~(\ref{e:CSigmaMeasurementStandard}) and
(\ref{e:CSigmaDetectionStandard}), are sufficient to enforce the
fundamental accuracy standards for any $\sigma$ in this range.  In the
extreme cases $\sigma=0$ and $\sigma=1$, ${\cal E}_{k,k'}(\sigma)$
reduces to one of the single $k$ waveform error measures and these
standards reduce to special cases of
Eqs.~(\ref{e:CkMeasurementStandard}) and
(\ref{e:CkDetectionStandard}).  For intermediate values $0<\sigma<1$
however, these time-domain waveform accuracy standards are completely
new and distinct.

\begin{figure}[t]
\centerline{\includegraphics[width=3in]{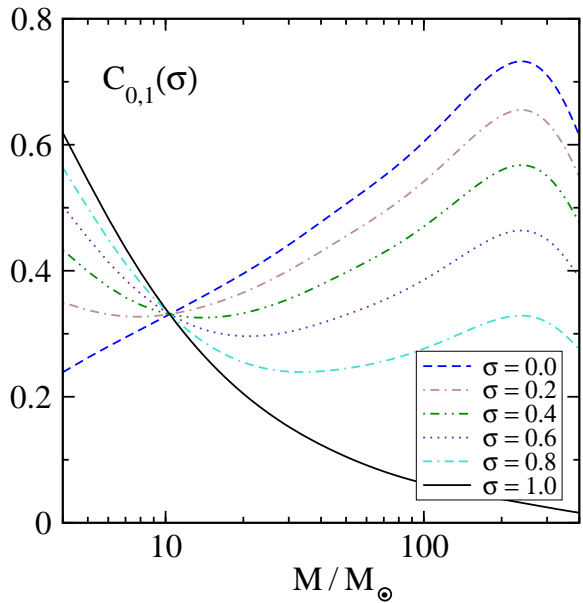}}
\caption{\label{f:CSigmaPlot} Curves show $
      C_{0,1}(\sigma)=\sqrt{\sigma
    C_0^2(M)+(1-\sigma)C^2_1(M)}$ as a function of mass for several
  values of the parameter $\sigma$.} 
\end{figure}
Figure~\ref{f:CSigmaPlot} illustrates $C_{0,1}(\sigma)=\sqrt{\sigma
  C_0^2+(1-\sigma)C^2_1}$ as a function of mass for the waveform and
noise curve used in the previous examples.  These graphs show that
$C_{0,1}(\sigma)$ reduces to $C_{0,1}(0)=C_1$ or $C_{0,1}(1)=C_0$ in
the limiting cases $\sigma=0$ or $\sigma=1$, respectively.  These
graphs also show that none of the $C_{0,1}(\sigma)$ is larger than one
of the limiting curves, $\sigma=0$ or $\sigma=1$, for any value of
$M$.  Consequently the new compound accuracy standards for
measurement, Eq.~(\ref{e:CSigmaMeasurementStandard}), are generally
more restrictive than the simpler single $k$ standard for measurement,
Eq.~(\ref{e:CkMeasurementStandard}).  Therefore, the new compound
accuracy standards are probably only useful for detection.

For detection, Eq.~(\ref{e:CSigmaDetectionStandard}) determines the
maximum allowed sizes of the error measures ${\cal E}_{k,k'}(\sigma)$.
Since these measures are independent of the mass of the waveform, it
follows that these detection standards will be satisfied for the full
range of masses only if they are satisfied for
$C_{k,k'}(\sigma)=\min_M C_{k,k'}(\sigma,M)$, where the
minimization is over the relevant range of masses.
Figure~\ref{f:CSigmaPlot} shows that these minima depend on the
parameter $\sigma$.  Perhaps the simplest of these compound error
measures is the one with $\sigma=0.5$, in which the two single $k$
error measures, ${\cal E}_k$ and ${\cal E}_{k'}$, are weighted
equally.  For the waveforms used in this example, $\min_M
C_{0,1}(0.5,M)\approx 0.314$, so the detection standard becomes ${\cal
  E}_{0,1}(0.5)\leq 0.314 \sqrt{2\epsilon_{\mathrm{max}}}$.  Since the
factor $0.314$ is a little smaller than the factor $0.331$ that
appears in Eq.~(\ref{e:C0C1DetectionStandard}), this particular
standard is considered a little more restrictive.  A deeper analysis
below however shows that this compound standard is more restrictive in
some cases but less restrictive in others compared to
Eq.~(\ref{e:C0C1DetectionStandard}).

There does exist, however, an unconditionally less restrictive
compound error standard for a particular optimal choice of the
parameter $\sigma=\sigma_{\mathrm{opt}}$.  Figure~\ref{f:CSigmaPlot}
shows that all the different $C_{0,1}(\sigma)$ curves intersect at the
single point $M=M_{\mathrm{int}}$ where
$C_0(M_{\mathrm{int}})=C_1(M_{\mathrm{int}})$.  Thus the largest of
the $\min_M C_{0,1}(\sigma,M)$ can never be larger than
$C_0(M_{\mathrm{int}})$.  The optimal value $\sigma_{\mathrm{opt}}$,
for which $\min_M
C_{0,1}(\sigma_{\mathrm{opt}},M)=C_0(M_{\mathrm{int}})$, can be
determined by enforcing the extremum condition, $d
C^2_{0,1}(\sigma,M)/dM=0$, at the point where $M=M_{\mathrm{int}}$.
The result is
\begin{eqnarray}
\sigma_{\mathrm{opt}}=\frac{dC_{1}^2}{dM}\left(\frac{dC_{1}^2}{dM}
-\frac{dC_0^2}{dM}\right)^{\!\!-1}\approx 0.284
\end{eqnarray}
for the waveform and noise curve example discussed in
Sec.~\ref{s:TimeDomainAccuracyStandards}. The optimal version of the
new compound time-domain accuracy standard for detection then becomes:
\begin{eqnarray}
{\cal E}_{0,1}(0.284)\lesssim
0.331\sqrt{2\epsilon_{\mathrm{max}}}.
\label{e:OptCSigmaDetectionStandard}
\end{eqnarray}
This new version of the detection standard uses the value
$C_{0,1}\approx 0.331$, so in this sense it is no more or less
restrictive than the previous standard,
Eq.~(\ref{e:C0C1DetectionStandard}).  However as the discussion below
shows, a larger portion of the two-dimensional space of error
measures, $({\cal E}_0, {\cal E}_1)$, is allowed using this new
standard.  So in this sense it is strictly less restrictive than the
detection accuracy standard given in
Eq.~(\ref{e:C0C1DetectionStandard}).

\begin{figure}[t]
\centerline{\includegraphics[width=3in]{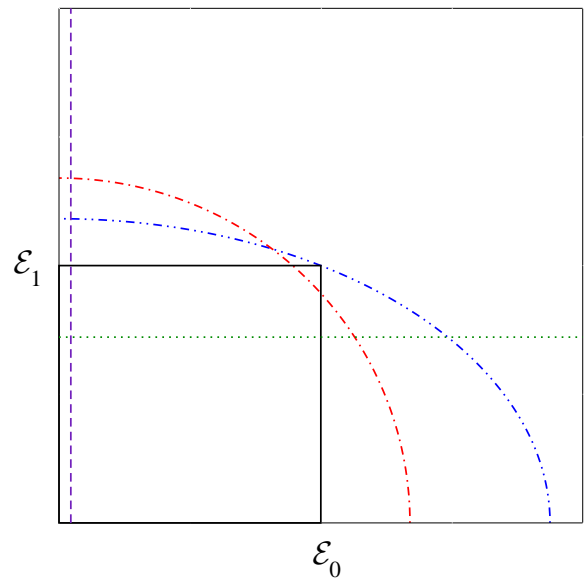}}
\caption{\label{f:Domain} Illustrates the allowed regions of the
  two-dimensional space of model waveform errors, $({\cal E}_0,{\cal
    E}_1)$, satisfying various versions of the accuracy standards
  discussed in the text.}
\end{figure} 
To help understand the relationship between these various new
standards, Fig.~\ref{f:Domain} illustrates the regions in the
two-dimensional space of error measures, $({\cal E}_0, {\cal E}_1)$,
allowed by each version.  The original time-domain standard,
Eq.~(\ref{e:C0DetectionStandard}) with $C_0\approx 0.016$, restricts
${\cal E}_0$ to the region left of the vertical dashed line in this
figure.  The simplest version of the new standard for detection, the
$k=1$ version of Eq.~(\ref{e:CkDetectionStandard}) with $C_1\approx
0.239$, restricts ${\cal E}_1$ to the region below the dotted
horizontal line.  A slightly more complicated standard,
Eq.~(\ref{e:C0C1DetectionStandard}) with $C_0=C_1=0.331$, restricts
both ${\cal E}_0$ and ${\cal E}_1$ 
to the region inside the solid-line
square.  The more complicated compound time-domain standards presented
in this Appendix restrict both ${\cal E}_0$ and ${\cal E}_1$ to the
region inside an ellipse defined in
Eq.~(\ref{e:CSigmaDetectionStandard}).  The simplest of these uses the
compound error measure ${\cal E}_{0,1}(\sigma)$ defined in
Eq.~(\ref{e:CompoundErrorNorm}) with $\sigma=0.5$.  This limits points
to the region inside the dash-dotted circle defined by
$C_{0,1}(0.5)\approx 0.314$.  The least restrictive standard of this
type uses the error measure ${\cal E}_{0,1}(0.284)$ with
$C_{0,1}(0.284)\approx 0.331$.  It limits points to the region inside
the dash-double-dotted ellipse shown in this figure.

Each of these versions of the waveform accuracy standard for detection
is sufficient to guarantee the original frequency-domain standards are
satisfied.  So any point, $({\cal E}_0, {\cal E}_1)$, in the
two-dimensional space of waveform errors, allowed by any of these
conditions is an acceptable standard satisfying point. The most
general accuracy standard of this type could be obtained by taking the
union of the allowed regions inside each of the ellipses defined by
Eq.~(\ref{e:CSigmaDetectionStandard}) with $C_{0,1}(\sigma)=\min_M
C_{0,1}(\sigma,M)$.  The boundary of this region qualitatively has the
shape of a hyperbola that asymptotes to the vertical dashed line and
the horizontal dotted line in Fig.~\ref{f:Domain}.  This region is
rather complicated to specify exactly, however, so this most general
version of these compound standards is probably impractical
for widespread use.

\break


\bibstyle{prd}
\bibliography{../References/References}
\end{document}